\documentclass[twocolumn]{aastex63}

\shorttitle{VLA Limits on Intermediate-Mass Black Holes}
\shortauthors{Wrobel \& Nyland}

\begin{document}

\title{VLA Limits on Intermediate-Mass Black Holes in 19 Massive Globular
  Clusters}

\author[0000-0001-9720-7398]{J. M. Wrobel}
\affiliation{National Radio Astronomy Observatory, P.O. Box O,
  Socorro, NM 87801, USA}

\author[0000-0003-1991-370X]{K. E. Nyland}
\affiliation{National Research Council, Resident at the U.S. Naval
  Research Laboratory, 4555 Overlook Avenue SW, Washington, DC 20375,
  USA}

\correspondingauthor{J. M. Wrobel}
\email{jwrobel@nrao.edu}

\accepted{2020 July 22 by ApJ}

\begin{abstract}

The NSF's Karl G.\ Jansky Very Large Array (VLA) was used at 3~cm to
search for accretion signatures from intermediate-mass black holes
(IMBHs) in 19 globular star clusters (GCs) in NGC\,3115, an early-type
galaxy at a distance of 9.4 Mpc.  The 19 have stellar masses
$M_{\star} \sim (1.1 - 2.7) \times 10^6~M_\odot$, with a mean
$\overline{M_{\star}} \sim 1.8 \times 10^6~M_\odot$.  None were
detected.  An IMBH accretion model was applied to the individual GCs
and their radio stack.  The radio-stacked GCs have an IMBH mass
$\overline{M_{\rm IMBH}} < 1.7 \times 10^5~M_\odot$ and mass fraction
$\overline{M_{\rm IMBH}} / \overline{M_{\star}} < 9.5\%$, with each
limit being uncertain by a factor of about 2.5.  The latter limit
contrasts with the extremes of some stripped nuclei, suggesting that
the set of stacked GCs in NGC\,3115 is not a set of such nuclei.  The
radio luminosities of the individual GCs correspond to X-ray
luminosities $L_{\rm X} < (3.3 - 10) \times 10^{38}$ erg~s$^{-1}$,
with a factor of about 2.5 uncertainty.  These limits predicted for
putative IMBHs in the GCs are consistent with extant {\em Chandra}
observations.  Finally, a simulated observation with a next-generation
VLA (ngVLA) demonstrates that accretion signatures from IMBHs in GCs
can be detected in a radio-only search, yet elude detection in an
X-ray-only search due to confusion from X-ray binaries in the GCs.
\end{abstract}

\section{MOTIVATION}\label{motivation}

Intermediate-mass black holes (IMBHs) are defined to have masses
$M_{\rm IMBH} \sim 10^2 - 10^5~M_\odot$, a range poorly explored
observationally.  Discovering them in the nuclei of dwarf galaxies or
in globular star clusters (GCs) would validate formation channels for
massive black hole (BH) seeds in the early universe \citep[for a
  review, see][]{gre20}.  Finding IMBHs in present-day GCs would also
inform predictions for gravitational wave (GW) and tidal disruption
(TD) events offset from galactic nuclei, and have broad implications
for the dynamical evolution of these compact stellar systems
\citep[for an overview, see][]{wro19}.

To search for IMBHs in GCs, one looks for evidence that the IMBHs are
influencing the properties of their GC hosts.  Within the Local Group,
a common approach is to scrutinize optical or infrared data for the
dynamical signatures of IMBHs on the orbits of stars in the GCs.
First suggested 50 years ago \citep{wyl70}, such sphere-of-influence
searches have a controversial history
\citep[e.g.,][]{noy08,noy10,and10,van10,bau17}, even resulting in
different IMBH masses when utilizing the orbits of stars or of pulsars
in the same GC \citep{per17,gie18}.  More fundamentally, a limitation
of dynamical searches is their susceptibility to measuring high
concentrations of stellar remnants rather than an IMBH
\citep[e.g.,][]{lut13,man19}.  It is thus important to develop
alternate ways to search for IMBHs in GCs.

One alternate approach capitalizes on decades of studies of accretion
signatures from both stellar-mass BHs and supermassive BHs \citep[for
  a review, see][]{fen16}.  This approach involves searching for
radio-synchrotron signatures of slow accretion onto putative IMBHs in
GCs, and was first suggested by \citet{mac04}.

Specifically, the semi-empirical model presented by \citet{str12} is
invoked to predict the mass of an IMBH that, if accreting slowly from
the tenuous gas supplied to the GC from its evolving stars, is
consistent with the synchrotron radio luminosity, $L_{\rm R}$, of the
GC. Summarizing from \citet{str12}, gas-capture is assumed to achieve
3\% of the Bondi rate \citep{pel05} for gas at a density of 0.2
particles~cm$^{-3}$ \citep{abb18} and a temperature of $10^4$~K.  It
is further assumed that accretion proceeds at less than 2\% of the
Eddington rate.  This establishes an inner advection-dominated
accretion flow with a predictable, persistent X-ray luminosity $L_{\rm
  X}$.  The empirical fundamental-plane of BH activity
\citep{mer03,fal04}, as refined by \citet{plo12}, is then used to
estimate the corresponding $L_{\rm R}$.  In this way, an observation
of $L_{\rm R}$ can lead to a constraint on $M_{\rm IMBH}$.  The radio
continuum emission from the IMBH is expected to be persistent,
flat-spectrum, jet-like but spatially unresolved, and located at or
near the dynamical center of the GC.  For context, a typical GC has a
half-starlight diameter of 5~pc \citep{bro06}.

Hereafter, this semi-empirical framework as implemented by
\citet{str12} is referred to as an IMBH accretion model.  Those
authors estimate that parameter uncertainties cause the $M_{\rm IMBH}$
associated with a given $L_{\rm R}$ to be uncertain by a factor of
about 2.5, dominantly from the gas-capture prescription.  Also,
\citet{str12} adopt a specific underlying Bondi flow that is
intermediate between the isothermal and adiabatic cases.  For a
general-case treatment otherwise consistent with \citet{str12},
including a general-case expression for $L_{\rm R}$, see Appendix C in
\citet{per17}.  The \citet{str12} framework is retained here for
consistency with prior work \citep{wro15,wro16,wro18,wro19}.

GCs with stellar masses ${M}_{\star} > 10^6~M_\odot$ are optimal
targets for radio searches, because contraints on their IMBH masses
can probe lower values for IMBH mass fractions $M_{\rm IMBH} /
{M}_{\star}$.  Early-type galaxies typically hold more GCs than other
galaxy types \citep[for a review, see][]{bro06}, increasing their
counts of massive GCs.  NGC\,3115 is among the nearest such galaxies:
it has hundreds of candidate GCs \citep{jen14,can15,can18} and a
distance of only 9.4~Mpc, where 1\arcsec = 45.6~pc \citep{bro14}.
Here, we report on radio observations of its massive GCs using the
NSF's Karl G.\ Jansky Very Large Array \citep[VLA;][]{per11}.
Section~2 describes the data and Section~3 explores their
implications.  Section~4 uses simulations to forecast the results of
substantially deeper observations with a next-generation VLA
\citep[ngVLA;][]{mur18}.  A summary and conclusions appear in
Section~5.

\section{DATA}\label{2}

\subsection{VLA Observations and Imaging}

\citet{jon19} observed NGC\,3115 with the VLA in its A configuration
under proposal VLA/14B-514, to investigate an optical
broad-emission-line nucleus claimed to be displaced from the
photometric center of the galaxy's stellar bulge.  Details of the
observing, calibration, and wide-field, wide-band imaging are as
described in \citet{jon19}, as is the validation of the VLA astrometry
by position-matching a background radio source with a X-ray
counterpart \citep{lin15} and a {\em ugi} counterpart
\citep{can15,can18}.  The VLA flux density scale is accurate to 3\%
\citep{perl17}.  The Common Astronomy Software Application (CASA)
package was used \citep{mcm07}.  About 1.4~hours of data were accrued
on NGC\,3115 at a wavelength of 3~cm, corresponding to a frequency of
10~GHz.  The bandwidth spanned 8 $-$ 12~GHz.

This work adopts an independent data analysis, also in CASA, which
improved the excision of radio frequency interference and yielded a
slightly lower rms noise in the image near the peak of the primary
beam.  After masking below a primary beam threshold of 20\%, the image
was corrected for the primary beam response and appears in Figure~1.
Its synthesized beam is elliptical with a major axis of 308~mas FWHM,
a minor axis of 168~mas FWHM, and an elongation position angle of 36
degrees.  Hereafter, the synthesized beam will be referred to by its
geometric-averaged resolution.  The galaxy's effective radius is
35\arcsec\, (Brodie et al.\ 2014) so Figure~1 samples out to about 5.7
effective radii.

\begin{figure}[t!]
\plotone{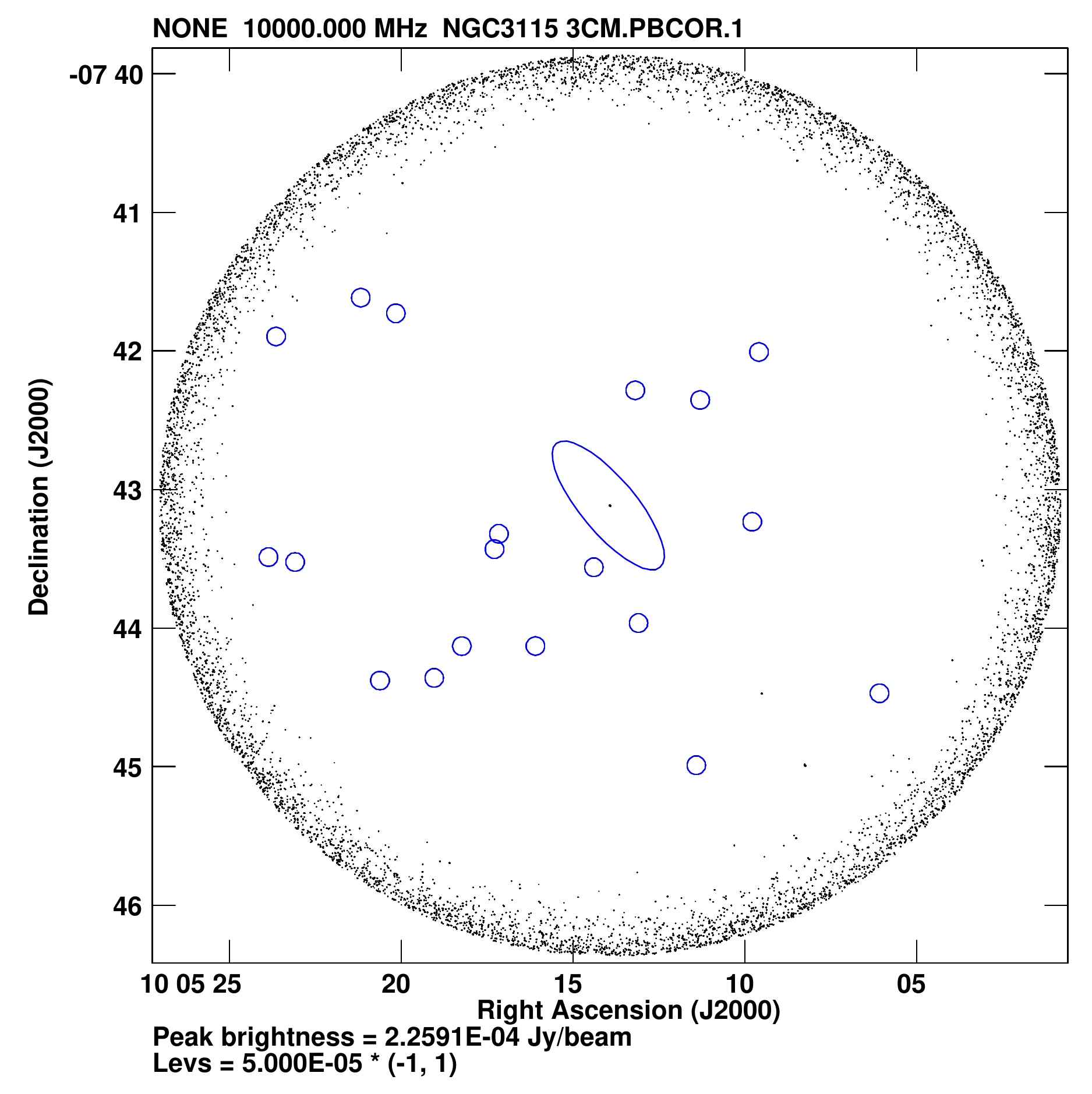}
\caption{VLA image of the Stokes $I\/$ emission at 3~cm centered near
  the radio nucleus of NGC\,3115 (Jones et al.\ 2019).  The image
  spans 6\farcm6 (18.1 kpc) on each axis and has a geometric-averaged
  resolution of 227~mas (10.4~pc).  The blue circles mark the
  positions of 19 massive GCs from Cantiello et al.\ (2018).  The blue
  ellipse marks the galaxy's effective radius (Brodie et al.\ 2014).
  It encloses a black dot at the galaxy's radio nucleus, which has
  traits consistent with those reported by Jones et al.\ (2019).  The
  apparent annulus of black dots is meant to illustrate the increased
  noise levels after correction for the primary beam
  attenuation.}\label{f1}
\end{figure}

The {\em g}-band magnitudes for the candidate GCs \citep{can18} were
converted to solar luminosities and then to stellar masses,
${M}_{\star}$, by assuming a typical GC mass-to-light ratio of 2.4
\citep{jor07}.  Massive, candidate GCs were then selected to have
${M}_{\star} > 10^6~M_\odot$ and positions within the
primary-beam-corrected region of Figure~1.  That region encompasses
only candidate GCs with negligible contamination from background {\em
  g}-band objects \citep{can18}, so the adjective ``candidate'' is
dropped hereafter.  To mitigate contamination from foreground stars,
\citet{can18} excluded {\em g}-band objects brighter than 19.5
magnitudes and thus set a stellar-mass cap of ${M}_{\star} \sim 3.8
\times 10^6~M_\odot$.  Nineteen massive GCs survived these selection
criteria.  Their properties are given in Table~1 and their positions
are marked in Figure~1.  The rms uncertainty for the positions is
200~mas and for the {\em g}-band photometry is typically 0.004~mag
\citep{can15,can18}.

% table 1 from t1f4.py yukon 3/27/2020
% idopt,g,starmass,rmsvla,lradio,bhmass,lxray,fractionlim
% added dropped zeros to tidy up columns
% added \\ to data rows
% data re-inserted 4/29/2020 as jumbled when caption squared up

\begin{deluxetable*}{rccccccccc}
\tablecolumns{10} \tablewidth{0pc} \tablecaption{19 Massive Globular
  Clusters in NGC\,3115} \tablehead{ 
  \colhead{ID} & \colhead{$\alpha_{\rm 2000}$} & 
  \colhead{$\delta_{\rm 2000}$} & \colhead{{\em g}} & 
  \colhead{$M_\star$} & \colhead{$\sigma_{\rm R}$} & 
  \colhead{4$\sigma_{\rm R}$ $L_{\rm R}$} & \colhead{$M_{\rm IMBH}$} &
  \colhead{$L_{\rm X}$} & \colhead{$M_{\rm IMBH}$/$M_\star$} \\
  \colhead{} & \colhead{(degrees)} & \colhead{(degrees)} &
  \colhead{(mag)} & \colhead{($10^6~M_\odot$)} & 
  \colhead{($\mu$Jy beam$^{-1}$)} & \colhead{($10^{34}$ erg s$^{-1}$)} &
  \colhead{($10^5~M_\odot$)} & \colhead{($10^{38}$ erg s$^{-1}$)} & 
  \colhead{(\%)} \\ 
  \colhead{(1)} & \colhead{(2)} & \colhead{(3)} & \colhead{(4)} &
  \colhead{(5)} & \colhead{(6)} & \colhead{(7)} & \colhead{(8)} & 
  \colhead{(9)} & \colhead{(10)}} 
\startdata 
   836 & 151.297577 & -7.749814 & 20.4450 & 1.58 & $<$6.52 & $<$2.76 & $<$3.22 & $<$5.52 & $<$20.4 \\ 
   852 & 151.275360 & -7.741160 & 20.7987 & 1.14 & $<$8.57 & $<$3.63 & $<$3.57 & $<$7.48 & $<$31.3 \\
   857 & 151.329361 & -7.739307 & 20.6693 & 1.28 & $<$6.04 & $<$2.56 & $<$3.13 & $<$5.07 & $<$24.4 \\
   858 & 151.335938 & -7.739635 & 20.0389 & 2.29 & $<$7.03 & $<$2.97 & $<$3.31 & $<$6.00 & $<$14.4 \\
   868 & 151.326019 & -7.735492 & 20.0891 & 2.19 & $<$5.14 & $<$2.17 & $<$2.95 & $<$4.24 & $<$13.5 \\
   876 & 151.317093 & -7.735482 & 20.7105 & 1.24 & $<$4.78 & $<$2.02 & $<$2.87 & $<$3.91 & $<$23.2 \\
   886 & 151.304581 & -7.732720 & 19.9794 & 2.42 & $<$4.27 & $<$1.81 & $<$2.76 & $<$3.45 & $<$11.4 \\
   902 & 151.309998 & -7.726009 & 20.7860 & 1.15 & $<$4.13 & $<$1.75 & $<$2.72 & $<$3.33 & $<$23.6 \\
   913 & 151.349472 & -7.724780 & 20.2986 & 1.81 & $<$9.61 & $<$4.07 & $<$3.72 & $<$8.49 & $<$20.6 \\
   916 & 151.346237 & -7.725379 & 20.7147 & 1.23 & $<$7.96 & $<$3.37 & $<$3.47 & $<$6.89 & $<$28.2 \\
   920 & 151.322052 & -7.723844 & 19.8486 & 2.73 & $<$4.24 & $<$1.79 & $<$2.75 & $<$3.42 & $<$10.1 \\
   932 & 151.321518 & -7.721987 & 20.6720 & 1.28 & $<$4.46 & $<$1.88 & $<$2.80 & $<$3.62 & $<$21.9 \\
   934 & 151.290802 & -7.720541 & 20.0403 & 2.29 & $<$4.47 & $<$1.89 & $<$2.80 & $<$3.63 & $<$12.2 \\
   990 & 151.297104 & -7.705904 & 20.7414 & 1.20 & $<$4.66 & $<$1.97 & $<$2.85 & $<$3.80 & $<$23.7 \\
   991 & 151.304977 & -7.704742 & 20.2613 & 1.87 & $<$4.37 & $<$1.85 & $<$2.78 & $<$3.54 & $<$14.9 \\
  1008 & 151.289978 & -7.700134 & 20.2837 & 1.83 & $<$5.34 & $<$2.26 & $<$2.99 & $<$4.42 & $<$16.4 \\
  1014 & 151.348541 & -7.698267 & 19.9248 & 2.55 & $<$11.4 & $<$4.83 & $<$3.97 & $<$10.3 & $<$15.6 \\
  1025 & 151.334015 & -7.695483 & 20.1705 & 2.03 & $<$6.37 & $<$2.69 & $<$3.20 & $<$5.38 & $<$15.7 \\
  1030 & 151.338272 & -7.693601 & 19.9196 & 2.56 & $<$8.74 & $<$3.70 & $<$3.59 & $<$7.64 & $<$14.0 \\
\enddata

\tablecomments{Columns (1)-(4) are from the Cantiello et al.\ (2018)
  catalog available at the CDS.  Details about Columns (5) and (6)
  appear in Section~2.1 and 2.2, respectively.  Columns (8)-(10) are
  predictions from the IMBH accretion model described in Section~1;
  entries for those columns are uncertain by a factor of about 2.5.}
\end{deluxetable*}

\subsection{VLA Cutouts}

For each of the 19 massive GCs, task {\tt subim} in NRAO's
Astronomical Image Processing System \citep[AIPS;][]{gre03} was used
to form a cutout spanning 20 times the geometric-averaged resolution
and centered on the tabulated position.  Figure~2 shows the VLA
cutouts for the 19 massive GCs.  Their rms noise levels range over
$\sigma_{\rm R} = (4.1 - 11)~\mu$Jy~beam$^{-1}$, as expected given the
the primary beam correction.  Based on the noise statistics among the
19 cutouts, a detection threshold of 4$\sigma_{\rm R}$ was adopted.
None of the 19 massive GCs are detected.

\begin{figure}[t!]
\plotone{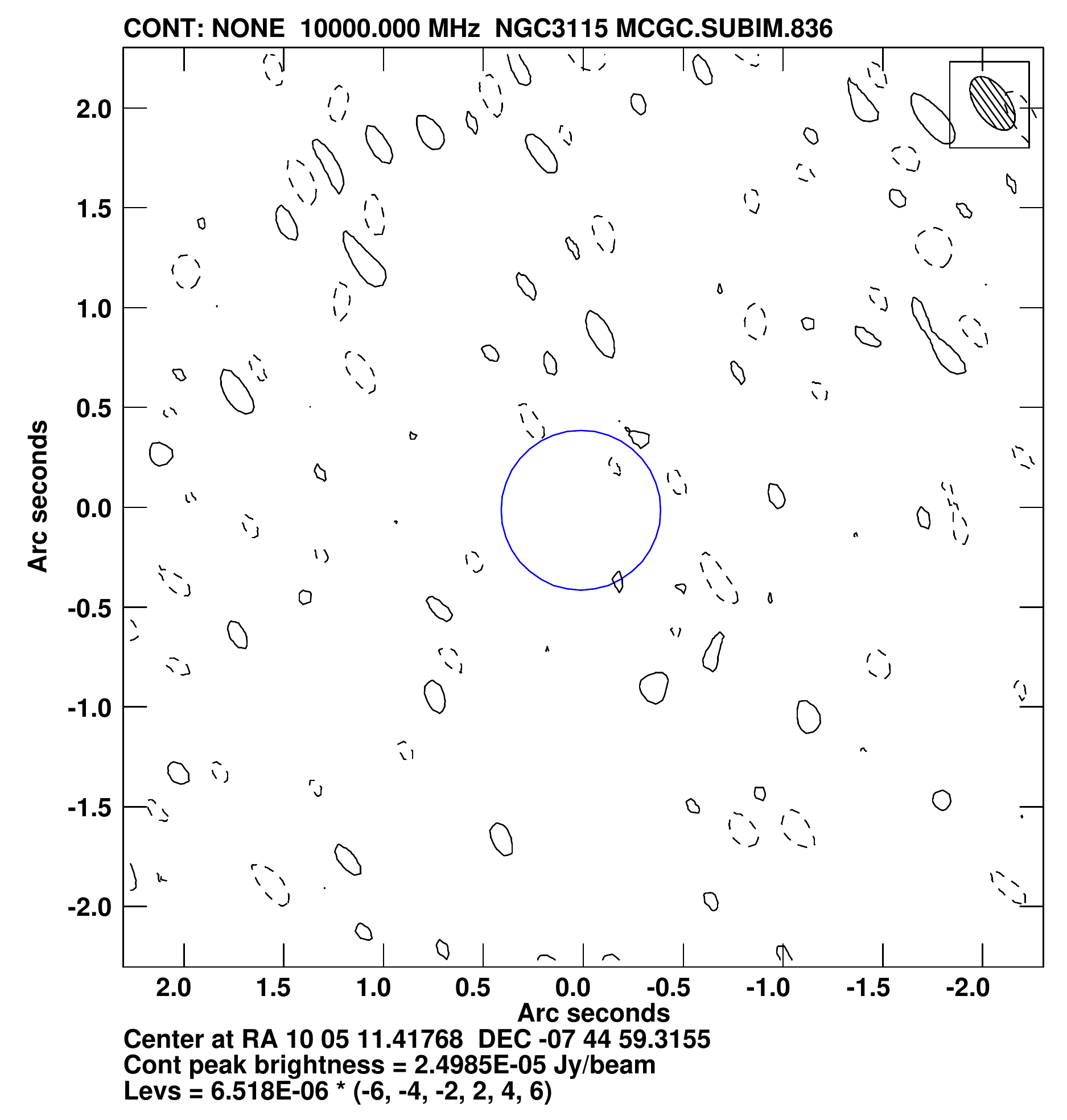}
\caption{Example VLA cutout of the Stokes $I\/$ emission at 3~cm
  centered on the {\em g}-band position of a massive GC labelled at
  the top as 836, its index ID in the Cantiello et al.\ (2018) catalog
  available at the CDS.  The cutout spans 4\farcs54 (207~pc) on each
  axis.  The hatched ellipse in the north-west corner shows the FWHM
  of the synthesized beam, which has a geometric average of 227~mas
  (10.4~pc).  Allowed contours are at -6, -4, -2, 2, 4, and 6 times
  the $\sigma_{\rm R}$ computed over the full cutout and given in the
  legend in units of Jy~beam$^{-1}$.  Dashed lines show negative
  contours and solid lines show positive ones.  The blue circle of
  diameter 800~mas (36~pc) conveys the 95\% positional uncertainty
  from Cantiello et al.\ (2015).  The VLA photometry seeks evidence
  for the accretion signature of a point-like IMBH in the center of
  the massive GC, which is not detected above the 4$\sigma_{\rm R}$
  level.  The complete figure set (19 images) is available in the
  online journal.}\label{f2}
\end{figure}

\subsection{VLA Stack}

Inferences from stacks are stronger if similar objects are involved.
The stellar masses of the 19 GCs appear to range over ${M}_{\star}
\sim (1.1 - 2.7) \times 10^6~M_\odot$ (Table~1).  This narrow span
helps strengthen inferences about other population-averaged traits of
the GCs from the stack of their radio upper limits shown in Figure 3.
The mean stellar mass of the GCs is $\overline{M_{\star}} \sim 1.8
\times 10^6~M_\odot$.

\begin{figure}[t!]
\plotone{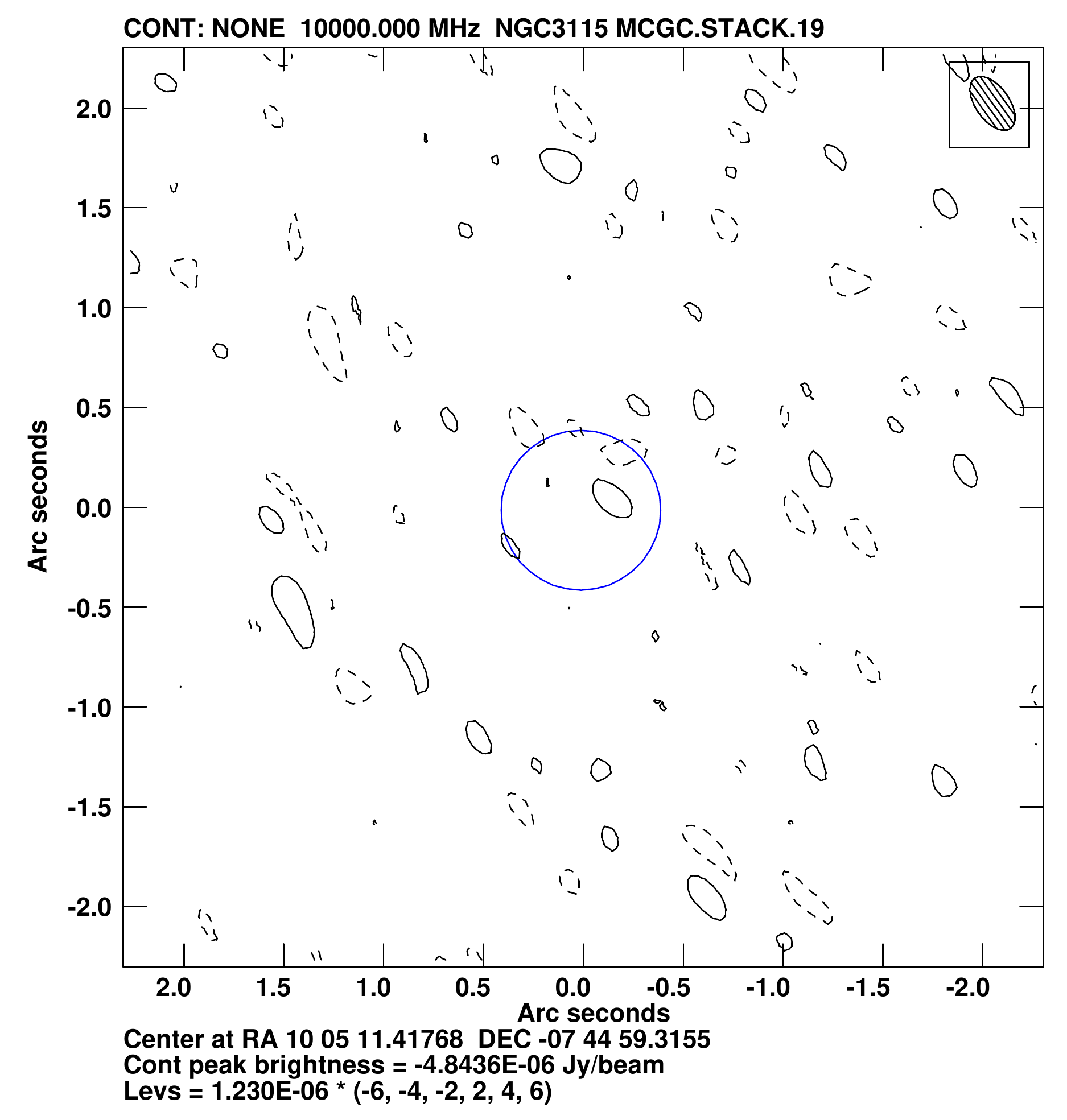}
\caption{Weighted-mean stack of the VLA cutouts of the 19 massive GCs,
  characterized by $\overline{\sigma_{\rm R}} = 1.23~\mu$Jy~ba$^{-1}$
  computed over the full cutout.  The hatched ellipse, blue circle and
  contouring scheme are the same as for Figure~2.  No emission is
  detected above 4$\overline{\sigma_{\rm R}} =
  4.92~\mu$Jy~beam$^{-1}$.}\label{f3}
\end{figure}

\section{IMPLICATIONS}\label{3}

\subsection{Applying the IMBH Accretion Model}

The upper limits on the radio flux densities for individual GCs
(Figure~2 and Table~1) and for their radio stack (Figure~3) were
converted to upper limits on the flat-spectrum radio luminosities.
Then, the IMBH accretion model was used to convert from radio
luminosities to putative IMBH masses, each uncertain by a factor of
about 2.5 \citep{str12}.

\subsubsection{The Individual GCs}

For each GC, its radio luminosity $L_{\rm R}$, IMBH mass $M_{\rm
  IMBH}$ and mass fraction $M_{\rm IMBH} / M_{\star}$ appear in
Table~1, and its $M_{\rm IMBH}$ is plotted in Figure~4 at its stellar
mass $M_{\star}$.  The constraints on the individual IMBH masses range
over $M_{\rm IMBH} < (2.7 - 4.0) \times 10^5~M_\odot$, whereas those
on the individual mass fractions range over $M_{\rm IMBH} / M_{\star}
< (10 - 31)\%$.  The black-hole mass fractions of some stripped
galactic nuclei are 40\% or more
\citep{set14,afa18,ahn17,ahn18,gre20}.  Such extreme fractions can be
ruled out for only a few individual GCs in NGC\,3115, given the factor
of about 2.5 uncertainty in their IMBH mass fractions.

\begin{figure*}[t!]
\plotone{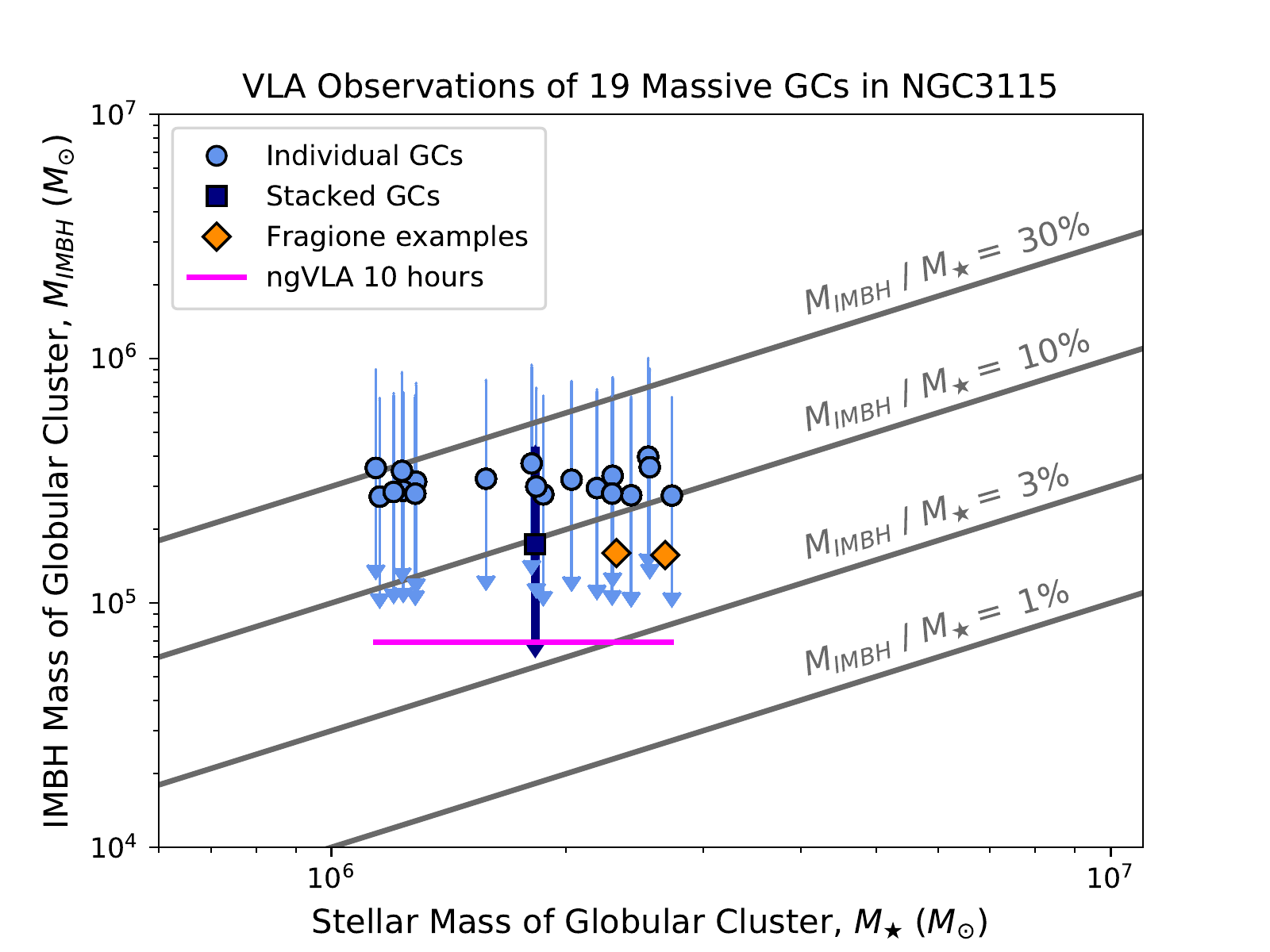}
\caption{Upper limit on the IMBH mass $M_{\rm IMBH}$ from the
  accretion model versus the stellar mass $M_{\star}$ of massive GCs
  in NGC\,3115.  Masses from the IMBH accretion model are uncertain by
  a factor of about 2.5, as conveyed by the error bars attached to the
  upper limits.  Sloping lines indicate mass fractions $M_{\rm
    IMBH}/M_{\star}$.  Example predictions from a semi-analytic model
  for GC evolution over cosmic time are shown (Fragione et al.\ 2018),
  as is the $M_{\rm IMBH}$ sensitivity from a simulated 10-hour
  observation with an ngVLA (Wrobel et al.\ 2018).}\label{f4}
\end{figure*}

The IMBH accretion model also yields the tabulated predictions for the
0.5 $-$ 10~keV X-ray luminosities $L_{\rm X}$, each uncertain by a
factor of about 2.5 \citep{str12}.  These predictions are consistent
with the very deep {\em Chandra} X-ray observations reported by
\citet{lin15}.  Specifically, X-ray sources 4, 23, 101, 106, 129, 188,
and 200 coincide with respective GC IDs 868, 932, 991, 1008, 934, 836,
and 852.  \citet{lin15} argue that those X-ray sources are low-mass
binaries with 0.5 $-$ 7~keV luminosities of about $10^{37-38}$
erg~s$^{-1}$, after adjusting for the closer distance to NGC\,3115
adopted in this work.  Those X-ray luminosities are compatible with
the $L_{\rm X}$ values for their respective GC hosts (Table~1).  The
balance of the GCs are undetected with {\em Chandra}, again at levels
compatible with their respective $L_{\rm X}$ values (Table~1).

\subsubsection{The Radio-Stacked GCs}

For the radio-stack of massive GCs in NGC\,3115, the radio luminosity
corresponds to $\overline{M_{\rm IMBH}} < 1.7 \times 10^5~M_\odot$ and
is plotted in Figure~4 at the stack's mean stellar mass
$\overline{M_{\star}} \sim 1.8 \times 10^6~M_\odot$.  The associated
mass fraction is $\overline{M_{\rm IMBH}} / \overline{M_{\star}} <
9.5\%$.

As mentioned in the previous subsection, some stripped galactic nuclei
have black-hole mass fractions of 40\% or more.  Such extreme mass
fractions contrast with the mass fraction for the radio-stacked GCs in
NGC\,3115, even given its factor of about 2.5 uncertainty.  This
suggests that the set of stacked GCs in NGC\,3115 is not a set of
highly-stripped galactic nuclei.  This finding is relevant to studies
of a galaxy's admixture of massive GCs and ultracompact dwarf
galaxies, from the perspectives of both simulations and observations
\citep[e.g.,][]{pfe16,nor19,dol20}.

For another early-type galaxy, NGC\,1023 at a distance of 11.1~Mpc,
similar radio-stacking results were obtained for 13 massive, old star
clusters with a mean stellar mass of $1.8 \times 10^6~M_\odot$: an
IMBH mass of less than $2.4 \times 10^5~M_\odot$ and a mass fraction
of less than 13\% \citep{wro15}.  Notably, these stacked limits on
IMBH masses already reach the mass regime targeted for dynamical
searches for black holes in stellar systems, as distant as NGC\,3115
and NGC\,1023, with next-generation near-infrared telescopes
\citep[e.g., Thirty Meter Telescope;][]{do14}.

\subsection{Investigating a GC Evolution Model}

The fate of primordial GCs, each assumed to host a central IMBH, was
first explored by \citet{hol08} and considerably elaborated on by
\citet{fra18}.  The latter authors modelled the evolution of the GCs
in a variety of host galaxies, as well as the IMBHs as they underwent
successive TD events and mergers with stellar-mass BHs in the GCs.
Summarizing from \citet{fra18}, such mergers generate successive GW
events, leading to predictions for the rates per volume of GW events
over time.  Meanwhile, the GC hosts lose mass due to stellar winds and
tidal stripping.  Among the primordial GCs that survived to the
present day, overall only a few percent retained their IMBHs and the
balance lost their IMBH when a GW recoil \citep{red89} ejected it from
its GC host.  Once ejected, the IMBHs are no longer able to foster GW
or TD events.  The more massive the GC, the higher the probability
that it will retain its IMBH to the present day.

\citet{lin20} reported a TD event from a massive star cluster at a
projected distance of 12.5~kpc from the center of a large early-type
galaxy.  This discovery is intriguing for two reasons.  First, the TD
event seems to match the \citet{fra18} predictions for the locations
of such events.  Second, the galaxy hosting the TD event and NGC\,3115
are both early-type galaxies.

With a total stellar mass of $8.5 \times 10^{10}~M_\odot$
\citep{for17} and an effective radius of 1.6~kpc \citep{bro14},
NGC\,3115 roughly matches a model for an early-type galaxy considered
by \citet{fra18}.  That model suggests that more than 100 GCs hosting
IMBHs could survive to the present day and reside at galactic radii of
up to 20~kpc.  Their projected radii would generally be lower, so
Figure~1 for NGC\,3115 samples radii relevant to the GC evolution
model.

\citet{fra18} provide two examples of how a star cluster's stellar
mass, IMBH mass, and mass fraction could evolve to yield a massive,
present-day GC.  Those examples, plotted in Figure~4, achieve
present-day values of $M_{\star} \sim (2.3 - 2.7) \times
10^6~M_\odot$, $M_{\rm IMBH} \sim 1.6 \times 10^5~M_\odot$, and
$M_{\rm IMBH} / M_{\star} \sim (6 - 7)\%$.  From Figure~4, it is clear
that neither the individual GCs nor the radio-stacked GCs in NGC\,3115
can provide useful constraints on these example IMBH masses and mass
fractions.  Such comparisons must await significant improvements to
the depth of the radio observations and/or the IMBH accretion model.
The former improvement area is addressed in Section~4.  Regarding the
latter area, the gas-capture prescription dominates the uncertainty in
the IMBH masses and mass fractions from the IMBH accretion model
(Section~1).  It would be desirable to replace that prescription with
the results of realistic simulations \citep[e.g.,][]{ina18} of gas
flows onto IMBHs in GCs.

\section{ngVLA SIMULATION}\label{4}

Deeper radio observations of the 19 massive GCs in NGC\,3115 would
improve the implications explored in Section~3.  The $M_{\rm IMBH}$
associated with a given $L_{\rm R}$ is uncertain by a factor of about
2.5 (Section~1), so it is reasonable to aim for an IMBH mass about
that factor below the $M_{\rm IMBH} \sim 1.6 \times 10^5~M_\odot$ of
the \citet{fra18} examples (Section~3.2).

Following \citet{wro18}, the IMBH accretion model was used to predict
the point-source flux density in the ngVLA band centered at 16.4~GHz
as a function of the mass $M_{\rm IMBH}$ of an IMBH in a GC at the
distance of NGC\,3115.  For selected point-source flux densities, the
ngVLA Sensitivity Calculator\footnote[3]{
  https://gitlab.nrao.edu/vrosero/ngvla-sensitivity-calculator}
provided estimates of the time to achieve $3\sigma$ detections with
the Main Array offering a synthesized beam of 100~mas FWHM (4.6~pc),
resembling that in Figures 1-3.  Picking an illustrative case, the
ngVLA reaches $3\sigma_{\rm ngVLA} = 0.25~\mu$Jy~beam$^{-1}$ after 10
hours on target, whereas observing with the VLA as in Section~2 would
require over 1000 hours on target.

This ngVLA detection threshold corresponds to $M_{\rm IMBH} \sim 6.9
\times 10^4~M_\odot$ and is conveyed in Figure~4.  Such an IMBH mass
is close to the desired factor below the \citet{fra18} examples
(Section~3.2), and would also enable exploration of mass fractions of
6\% or less among individual GCs in NGC\,3115, thus aiding comparisons
with stripped galactic nuclei (Section~3.1).  Moreover, this ngVLA
threshold corresponds to $L_{\rm X} \sim 5.5 \times 10^{36}$
erg~s$^{-1}$.  Although uncertain by a factor of about 2.5, this value
is below or near the X-ray luminosities of the low-mass binaries in
GCs in NGC\,3115 (Section~3.1).  This illustrates how an X-ray-only
search for the accretion signatures of IMBHs in GCs could be hindered
by confusion from X-ray binaries in GCs, an issue further developed in
\citet{wro18}.

\section{SUMMARY AND CONCLUSIONS}\label{5}

The VLA was used at 3~cm to search for accretion signatures from
putative IMBHs in 19 GCs in the early-type galaxy NGC\,3115.  The
stellar masses of the 19 have a range ${M}_{\star} \sim (1.1 - 2.7)
\times 10^6~M_\odot$ and a mean $\overline{M_{\star}} \sim 1.8 \times
10^6~M_\odot$.  None of these massive GCs were detected, leading to
the following findings:

\begin{itemize}

\item An IMBH accretion model was applied to the radio-stacked GCs in
  NGC\,3115, resulting in an IMBH mass $\overline{M_{\rm IMBH}} < 1.7
  \times 10^5~M_\odot$ and mass fraction $\overline{M_{\rm IMBH}} /
  \overline{M_{\star}} < 9.5\%$.  The latter limit contrasts with the
  extremes of some stripped nuclei, suggesting that the set of stacked
  GCs in NGC\,3115 is not a set of such nuclei.  Neither limit was
  accurate enough for useful comparisons with the present-day
  predictions of a recent semi-analytical model for GC evolution.
  Such comparisons must await significant improvements to the IMBH
  accretion model and/or the depths of the radio observations.

\item Within the context of the IMBH accretion model, the radio
  luminosities of the individual GCs in NGC\,3115 correspond to X-ray
  luminosities $L_{\rm X} < (3.3 - 10) \times 10^{38}$ erg s$^{-1}$.
  These predicted limits for IMBHs in GCs are consistent with existing
  {\em Chandra} observations.

\item An ngVLA simulation of NGC\,3115 showed that accretion
  signatures from IMBHs in GCs can be detected in a radio-only search,
  yet escape detection in an X-ray-only search because of confusion
  from X-ray binaries in the GCs.

\end{itemize}

\acknowledgments The authors thank Dr.\ Tom Maccarone for discussions
and the reviewer for a timely and helpful report.  The National Radio
Astronomy Observatory is a facility of the National Science
Foundation, operated under cooperative agreement by Associated
Universities, Inc.  The Next Generation Very Large Array (ngVLA) is a
design and development project of the National Science Foundation
operated under cooperative agreement by Associated Universities, Inc.
Basic research in radio astronomy at the U.S. Naval Research
Laboratory is supported by 6.1 Base Funding.

\facility{VLA}

\software{AIPS (Greisen 2003), astropy (The Astropy Collaboration 2018),
CASA (McMullin et al.\ 2007)}

\clearpage

% For papers with more than five authors, the last name and initials
% of the first three authors only should be listed, followed by a
% comma and ``et al.''

\end{document}